\documentclass[aps,prd,10pt,twocolumn,nofootinbib]{revtex4}

\usepackage{epsfig}
\usepackage{bm}
\usepackage{latexsym}
\usepackage{natbib}
\usepackage{url}
\usepackage{dcolumn}
\usepackage{color}
\usepackage{amsfonts,amssymb,amsmath}
\usepackage{graphicx,epsfig}
\usepackage{psfrag}
\usepackage{subfigure}
\usepackage{hyperref}
\hypersetup{colorlinks=true}
\usepackage{mathtools}
\usepackage{enumitem}
\usepackage{float}
\usepackage[dvipsnames]{xcolor}

\def\beq{\begin{equation}}
\def\eeq{\end{equation}}

\begin{document}

\title{Swampland, Axions and Minimal Warm Inflation}

\author{
Suratna Das$^{a,}$\footnote{suratna@iitk.ac.in}, 
Gaurav Goswami$^{b,}$\footnote{gaurav.goswami@ahduni.edu.in} and
Chethan Krishnan$^{c,}$\footnote{chethan.krishnan@gmail.com}
}

\affiliation{
$^a$ Department of Physics, Indian Institute of Technology, Kanpur, Kanpur 208016, India\\
$^b$ School of Engineering and Applied Science, Ahmedabad University, Ahmedabad 380009, India\\
$^c$ Center for High Energy Physics, Indian Institute of Science, Bangalore 560012, India
}

\begin{abstract}
Warm inflation has been noted previously as a possible way to implement inflationary models compatible with the dS swampland bounds. But often in these discussions the heat bath dynamics is kept largely unspecified. We point out that the recently introduced Minimal Warm Inflation of \cite{Berghaus:2019whh}, where an axionic coupling of the inflaton leads to an explicit model for the thermal bath, yields models of inflation that can easily fit cosmological observations while satisfying de Sitter swampland bounds, as well as the swampland distance bound and trans-Planckian censorship.  
 
\end{abstract}

\maketitle

\section{Introduction}

Which low energy effective theories can arise from a UV complete theory of quantum gravity (such as string theory), is a question of both theoretical and phenomenological  interest \cite{Brennan:2017rbf, Danielsson:2018ztv, Palti:2019pca, Roupec:2018mbn}. In particular, inspired by the difficulty of realizing inflation and/or de Sitter vacua in string theory, it has recently been conjectured that scalar potentials whose potential slow roll parameters are small, cannot be realised in (asymptotic regimes of) string theory \cite{Obied:2018sgi,Garg:2018reu,Ooguri:2018wrx}.
Since conventional models of cold inflation require small potential slow roll parameters, if one wants to have inflation in such regimes, one must explore alternative models. One simple way to achieve sufficient amounts of inflation, even for steep potentials, is to employ the ``warm inflation" mechanism in the strongly dissipative regime (see \cite{Berera:1995wh,Berera:1995ie} for some early papers and \cite{Berera:2008ar,Rangarajan:2018tte} for review). In the warm inflation paradigm, the inflaton loses its energy to a thermal bath. Its utility for swampland purposes has been noted previously \cite{Das:2018hqy, Motaharfar:2018zyb, Das:2018rpg, Kamali:2019ppi, Das:2019hto, Kamali:2019xnt,Berea:2019tcc}.

On the other hand, while warm inflation has been studied for a long time as a possibility, realising it in concrete models has been  a challenge (see e.g. the discussion in \cite{Bastero-Gil:2016qru,Bastero-Gil:2019gao} and references therein). In particular, any endeavour to realise warm inflation in a strongly dissipative regime has difficulties because the strong dissipation typically destabilizes the inflationary potential. 

Very recently however, a class of concrete models (``Minimal Warm Inflation") that realize warm inflation in the strongly dissipative regime, have been put forward \cite{Berghaus:2019whh} (see also \cite{Bastero-Gil:2019gao}). Minimal Warm Inflation gives the inflaton an axionic coupling to non-Abelian gauge fields (much like the Abelian mechanism studied in \cite{Anber:2009ua,Anber:2012du}). 
This provides a very simple and possibly viable model of the thermal bath. Since the inflaton is an axion, its shift symmetry will protect it from any perturbative backreaction and hence from acquiring a large thermal mass. On the other hand, because it is coupled to the gauge field and since at sufficiently high temperature there are sphaleron transitions between gauge vacua, there is friction. The corresponding axion friction coefficient, $\Upsilon$, turns out to be \cite{Berghaus:2019whh} (see e.g. section 9.5 of \cite{Laine:2016hma} and also \cite{Gross:1980br,Arnold:1987mh,Moore:2010jd,Frison:2016vuc})
\begin{equation} \label{eq:upsi}
\Upsilon (T) = \frac{\Gamma_{sp}(T)}{2 f^2 T} = \kappa (\alpha_g, N_c, N_f) \alpha_g^5 \frac{T^3}{f^2} \; ,
\end{equation}
where, $T$ is the temperature of the bath, $\Gamma_{sp}(T)$ is the sphaleron rate, $f$ is the axion decay constant, $\alpha_g = g^2/(4\pi)$, $g$ being the Yang-Mills gauge coupling, and $\kappa$ is a dimensionless quantity which depends on the dimension of the gauge group ($N_c$), the representation of fermions ($N_f$) if any, and on the gauge coupling.
In addition to this, the axion has a UV potential that is responsible for inflation, which (it is hoped) softly breaks the shift symmetry without causing too much backreaction \cite{Berghaus:2019whh}. 

Since axions and gauge fields are ubiquitous in string theory, the mechanism of \cite{Berghaus:2019whh} has ingredients which may be realizable in string theory. But, for many stringy solutions (near the boundary of the landscape), we also know that the scalar potential violates potential slow-roll \cite{Obied:2018sgi,Garg:2018reu,Ooguri:2018wrx} as was first noted in the example of \cite{Hertzberg:2007wc}. This raises the following question: could the ingredients used in \cite{Berghaus:2019whh}, which lead to inflation in the specific models studied there, lead to inflation when one is dealing with potentials compatible with the string landscape? 

In this short note, we would like to point out that a simple model in which the inflaton is an axion and its UV potential can be made compatible with the swampland constraints, is a viable model to achieve warm inflation in the strongly dissipative regime. We will show that (a) CMB observational constraints are easily satisfied, (b) the dS swampland bounds are satisfied, (c) the field excursion can be sub-Planckian so that the requirement of swampland distance conjecture \cite{Ooguri:2006in,Klaewer:2016kiy} is satisfied, and (d) the energy scale of inflation is low enough so that the recently proposed trans-Planckian censorship conjecture \cite{Bedroya:2019snp,Bedroya:2019tba} holds
\footnote{In this context, it is important to note some of the other recent attempts to come up with inflation models consistent with dS swampland conjecture \cite{Brahma:2018hrd,Ashoorioon:2018sqb} and trans-Planckian censorship conjecture \cite{Cai:2019hge,Mizuno:2019bxy,Brahma:2019unn,Dhuria:2019oyf,Schmitz:2019uti}.}. 
In the following, we will elaborate on these claims. 

In this paper, our primary focus will be on showing that enough inflation to simultaneously satisfy observational data and swampland constraints is possible. In particular, we will not concern ourselves with ending inflation, and will assume the existence of suitable mechanisms for accomplishing it. In a concluding section, we will comment on what it takes for our scenario to be turned into a full cosmological model.  
 
\section{Minimal Warm Inflation and the swampland}

\subsection{Equations and approximations}

For warm inflation, at the background level, one is interested in the dynamics of the homogeneous inflaton field $\phi(t)$ and temperature of the bath $T(t)$. The evolution equations are
\begin{eqnarray}
 {\ddot \phi} + 3 H {\dot \phi} + \Upsilon(T) {\dot \phi} + V'(\phi) &=& 0 \; , \label{eq:phi}  \\
H^2- \frac{1}{3 M_{pl}^2} \left( \rho_R + \frac{{\dot \phi}^2}{2} + V(\phi) \right) &=& 0 \; , \label{eq:Hubb} \\ 
 \dot{\rho}_R + 4H \rho_R - \Upsilon(T) {\dot \phi}^2 &=& 0 \; , \label{eq:rad}
\end{eqnarray}
where $\Upsilon(T)$ is the axion friction coefficient, which, for our purpose, is given by Eq (\ref{eq:upsi}) and $\rho_R = {\tilde g}_* T^4$ where 
${\tilde g}_* = \frac{\pi^2 g_*}{30}$, all the other symbols have their usual meaning. 
It is useful to work with the dimensionless quantity $Q$ defined by
\begin{equation}  \label{eq:Q}
Q = \frac{\Upsilon}{3H} \; .
\end{equation}
In the following, we follow the convention of the literature on cold inflation and define the potential slow roll parameters in the usual following way,
\begin{equation}
\epsilon_V = (M_{pl}^2/2) (V'/V)^2 \; ,~  \eta_V = M_{pl}^2 (V''/V) \; .
\end{equation}
In the literature on warm inflation, it is usual to define another set of slow roll parameters, for which we use the following notation:
\begin{eqnarray}
 \epsilon_w &=& \frac{\epsilon_V}{(1+Q)} \; , \label{eq:epsw} \\
 \eta_w &=& \frac{\eta_V}{(1+Q)} \; , \label{eq:etaw}
\end{eqnarray} 
in addition, one can have the usual Hubble slow-roll parameters. 
Before proceeding, let us note the following important points:
\begin{enumerate}[label=(\alph*)] 
 \item During warm inflation, friction due to the thermal bath ensures that the inflaton slow rolls even when the potential is steep, this means that the ${\ddot \phi}$ term in Eq (\ref{eq:phi}) can be ignored,
 \item We want the Universe to inflate, so we need $V(\phi)$ to be larger than ${\dot \phi}^2/2$ and $\rho_R$ (energy density of radiation). Thus, in Eq (\ref{eq:Hubb}), the only term in bracket which is relevant is $V(\phi)$,
 \item We want to deal with warm inflation in strongly dissipative regime, this corresponds to $3 H \ll \Upsilon(T)$ in Eq (\ref{eq:phi}) i.e. the condition 
\begin{equation}
 Q \gg1 \; ,
\end{equation}
 \item It can be shown that, when $\epsilon_w \ll 1$ and $\eta_w \ll 1$, the first term in Eq (\ref{eq:rad}), i.e. $\dot{\rho}_R$ can be ignored. Notice that in strongly dissipative regime with  $Q \gg1$, one can have  $\epsilon_w$ and $\eta_w$ too small even if $\epsilon_V$ and $\eta_V$ are ${\cal O}(1)$; 
 \item Finally, let us note that warm inflation requires that $T>H$ and when this does not hold good, we are dealing with cold inflation. 
\end{enumerate}
With all the above approximations, Eq (\ref{eq:phi}), Eq (\ref{eq:Hubb}) and Eq ({\ref{eq:rad}) take the form:
\begin{eqnarray}
 {\dot \phi} &\approx& \frac{-V'(\phi)}{3 H (1+Q)} \; , \label{eq:phi_a}  \\
H^2 &\approx& \frac{V(\phi)}{3 M_{pl}^2} \; , \label{eq:Hubb_a} \\ 
\rho_R &\approx& \frac{3 Q {\dot \phi}^2}{4} \; , \label{eq:rad_a}
\end{eqnarray}
notice that we are {\it not} assuming that $\epsilon_V$ and $\eta_V$ are small.

\subsection{Constraints on potential}

\subsubsection{de Sitter swampland constraint}

The de Sitter swampland bounds \cite{Obied:2018sgi,Garg:2018reu,Ooguri:2018wrx} (see also \cite{Zaz, Andriot}) dictate that at least one of the potential slow roll parameters among $\epsilon_V$ or $\eta_V$ must be an ${\cal O}(1)$ number in Planck units. More specifically,
\begin{equation} \label{eq:ds-swampland-condition}
 {\rm either}~\epsilon_V \gtrsim {\cal O}(1) \; ,~{\rm or}  \ ~  \eta_V \lesssim -{\cal O}(1) \; .
\end{equation}

\subsubsection{Constraint from swampland distance conjecture}

Swampland distance conjecture \cite{Ooguri:2006in,Klaewer:2016kiy} states that as we explore distances comparable to $M_{pl}$ in scalar field space, towers of states become exponentially light. Thus, a potential obtained from low energy effective field theory can only be trustworthy for sub-Planckian field excursions. 

\subsubsection{Trans-Planckian censorship}

The Trans-Planckian Censorship Conjecture \cite{Bedroya:2019snp,Bedroya:2019tba} states that the cosmological evolution in effective theories consistent with quantum gravity must be such that quantum fluctuations at sub-Planckian length scales must never become classical. This requirement imposes strong constraints on cosmic inflation. In particular, it is argued in \cite{Bedroya:2019tba} that this will imply that the potential energy during inflation, $V$, must satisfy the inequality (see also \cite{Berea:2019tcc})
\begin{equation}
 V^{1/4} < 3 \times 10^{-10} M_{pl} \; .
\end{equation}

\subsubsection{Observational constraints}

Finally, we have the observational constraints. According to 2018 Planck \cite{Akrami:2018odb} TT,TE,EE+lowE+lensing data, 
the amplitude of scalar perturbations is known to be $A_s = 2.0989^{+0.0296}_{-0.0292} \times 10^{-9}$, the scalar spectral index is measured to be
$n_s = 0.9649\pm0.0042$ at $68\%$ C.L. while the tensor to scalar ratio $ r_{0.002} < 0.10$ at 95\% C.L. (this corresponds to the pivot scale of $k_* = 0.002 {\rm Mpc}^{-1}$). On the other hand, when one combines 2018 Planck with BICEP2/Keck Array 2014 B-mode polarization data i.e. TT,TE,EE+lowE+lensing+BK14, one finds $r_{0.002} < 0.064$ (95\% C.L.).

\subsection{General method of analysis}

Before proceeding, we rewrite Eq (\ref{eq:upsi}), in the form $\Upsilon(T) = {\tilde c}~T^3$, where, 
\begin{equation} \label{def:ctilde}
 M_{pl}^2 ~{\tilde c} = \frac{\kappa \alpha_g^5}{(f/M_{pl})^2} \; .
\end{equation}
At the most basic level, for background evolution, we are dealing with the following basic quantities: $\phi(t), {\dot \phi}(t)$ and $T(t)$. In addition, we have equations which determine some of the other quantities of interest in terms of these basic quantities e.g. $\Upsilon(T) = {\tilde c}~T^3$, $\rho_R (T) = {\tilde g}_* T^4$, 
$H^2 (\phi, {\dot \phi}, T) =  \frac{1}{3 M_{pl}^2} \left( \rho_R + \frac{{\dot \phi}^2}{2} + V(\phi) \right)$ and 
$Q = \frac{\Upsilon(T)}{3 H}$.
Furthermore, we have Eq (\ref{eq:phi}), Eq (\ref{eq:Hubb}) and Eq (\ref{eq:rad}) which determine the evolution of these quantities.

The free parameters available are ${\tilde c}, {\tilde g}_*$, parameters in $V(\phi)$, $N_{\rm cmb}$ (the number of e-foldings of inflation after the pivot scale crossed the Hubble radius during inflation), 
and the initial conditions are $\phi(t_i), {\dot \phi}(t_i), T(t_i)$. 
In addition, we have quantities such as $\phi_{\rm end}$ (the inflaton field value at the end of inflation) and $\phi_*$ (the inflaton field value when the pivot scale crossed the Hubble radius during inflation).
We also have observational constraints e.g. $A_s$, $n_s$, $r$ and theoretical constraints such as refined de Sitter Swampland conjecture, Trans-Planckian Censorship Conjecture and Swampland distance conjecture.

Given all of this, the key question one might wish to answer could be: for a given potential, what should be the initial conditions and values of parameters, such that we satisfy all observational constraints and as many theoretical constraints as possible?

In the rest of this paper, we shall answer this question analytically as well as numerically.
When we use analytical arguments, we shall work with the approximate equations: Eq (\ref{eq:phi_a}), Eq (\ref{eq:Hubb_a}) and Eq (\ref{eq:rad_a}). For numerical work, we shall work with Eq (\ref{eq:phi}), Eq (\ref{eq:Hubb}) and Eq (\ref{eq:rad}).

For analytical arguments, we could define $\phi_{\rm end}$ by the requirement that 
\begin{equation}
\epsilon_w = \frac{\epsilon_V}{(1+Q)} = 1 \; ,\label{endinf}
\end{equation}
while when doing numerical work, we could find exact field value at which the Universe starts decelerating.
In general, one expects $\epsilon_w$ to be dependent on $\phi, {\dot \phi}, T$, but 
we have approximate expressions for $Q(\phi)$ and $T(\phi)$ which can be derived inserting the set of equations given in 
Eq (\ref{eq:phi_a}),  (\ref{eq:Hubb_a}) and (\ref{eq:rad_a}) (with $\rho_R=(\pi^2/30)g_*T^4$) into the relation $Q\equiv \Upsilon/3H={\tilde c}~ T^3/3H$. Then, using Eqs (\ref{eq:phi_a}),  (\ref{eq:Hubb_a}) and (\ref{eq:rad_a}), we get the forms of $Q$ and $T$ 
as (see also \cite{Berghaus:2019whh})
\begin{eqnarray}
Q^7 &\approx& \frac{1}{576} \frac{(M_{pl}^2 {\tilde c})^4}{{\tilde g}_*^3} ~M_{pl}^2~ \frac{V'(\phi)^6}{V(\phi)^5} \equiv \tilde C\frac{V'(\phi)^6}{V(\phi)^5} \; ,  \label{eq:Q} \\
T^7&\approx&\frac{\sqrt{3}}{4}\frac{1}{(M_{pl}^2\tilde c) \tilde g_*}M_{pl}^3\frac{V'^2(\phi)}{V^{1/2}(\phi)}\:,
\label{eq:QandT}
\end{eqnarray} 
these expressions are only applicable when $Q \gg 1$. Notice that in Eq (\ref{eq:Q}), we have introduced a variable named $\tilde C$ which should not be confused with ${\tilde c}$ defined by Eq (\ref{def:ctilde}).
Now, one can find an analytical condition for $\phi_{\rm end}$ by using $\epsilon_V=1+Q$, which yields 
\begin{eqnarray}
\left.\frac{V'^{8/7}}{V^{9/7}}\right|_{\phi_{\rm end}}=\frac{2}{M_{pl}^2}\tilde C^{1/7}.
\label{phi-end}
\end{eqnarray}
Now, once we have $\phi_{\rm end}$ we can use
\begin{equation} \label{eq:ncmb-warm}
N_{\rm cmb} = - \int^{\phi_{\rm end}}_{\phi_*} \frac{d \phi}{M_{pl}} ~\frac{1+Q(\phi)}{M_{pl}}~ \frac{V(\phi)}{V'(\phi)} \; ,
\end{equation}
to find $\phi_*$ for a chosen value of $N_{\rm cmb}$. One can then determine $Q_*$, $T_*$ etc.
At this stage, one can look for the parameters which lead to the correct values of $A_s$ and $n_s$ and other requirements.
In the approximation that $Q \gg 1$, one has (see e.g. Eq (75) of \cite{Graham:2009bf}, the discussion around Eq (4.18) of \cite{BasteroGil:2011xd} and \cite{Berghaus:2019whh})
\begin{eqnarray}
A_s &=& \frac{1}{4 \pi^{3/2}} \frac{T Q^{5/2}}{M_{pl}^5} \left( \frac{Q}{Q_3} \right)^9 \frac{V^{5/2}}{(V')^2} \; , \label{eq:As}\\
n_s &=& 1 + \frac{3}{7} \frac{27 \epsilon_V - 19 \eta_V}{1+Q} \; , \label{eq:ns} \\
r&\approx& \frac{1}{\sqrt{3\pi}}\frac{16\epsilon_V}{Q^{3/2}}\frac HT\left(\frac{Q_3}{Q}\right)^9\label{eq:r}.
\end{eqnarray}
where $Q_3\sim7.3$.

Finally, let's also find how $Q$ evolves with $N$. To see that we will derive a more accurate relation for $Q$. Inserting $\Upsilon=\tilde c~ T^3$ and $\rho_R=\tilde g_*T^4$ in Eq (\ref{eq:rad_a}), we get $T=(\tilde c/4\tilde g_*)(\dot\phi^2/H)$. Inserting this into $Q=\tilde c T^3/3H$ and using set of equations (\ref{eq:phi_a}), (\ref{eq:Hubb_a}) and (\ref{eq:rad_a}), we get
\begin{eqnarray}
(1+Q)^6Q=\tilde C\frac{V'^6}{V^5}.
\end{eqnarray}
By taking log on both sides and then taking a derivative wrt $N$ would yield 
\begin{eqnarray}
\frac1Q\frac{dQ}{dN}=\frac{10\epsilon_V-6\eta_V}{1+7Q} \; .
\label{Q-evo}
\end{eqnarray}
This gives the rate of change of $Q$ as inflation proceeds.

\section{Warm inflation with run-away potentials}

In this section, we will carefully analyse inflationary predictions of scalar potentials consistent with de Sitter Swampland conjecture. By the end of this section, we shall present a model of inflation which has the following features:
(a) its scalar potential is consistent with de Sitter Swampland conjecture (i.e. it has a steep potential),
(b) the inflaton field excursion required to achieve sufficient number of $e$-foldings of inflation is sub-Planckian (as expected from Swampland distance conjecture),
(c) the energy scale of inflation (and the corresponding number of $e$-folds) is consistent with the Trans-Planckian Censorship Conjecture, 
(d) it is based on warm inflation realised in a strongly dissipative regime,
(e) the model is a Minimal Warm Inflation model and thus, there is a clear understanding of the thermal bath in terms of axionic couplings of the inflaton, and, finally,
(f) the model is consistent with cosmological observations.
Before proceeding, one must note that in this model, there will be no graceful exit from inflation and we shall not address this issue in the present paper. 

\subsection{Runaway potential of the form $V = V_0 e^{-\alpha \frac{\phi}{M_{pl}}}$}
\label{sec:oldpot}

Let us first begin to analyse potentials of the form 
\begin{equation} \label{eq:oldpot}
V = V_0 e^{-\alpha \frac{\phi}{M_{pl}}} \; ,
\end{equation}
which are expected to be consistent with dS swampland conjecture.
This is because, for such a potential, one finds that $\epsilon_V = \alpha^2/2$ while $\eta_V = \alpha^2$ so that $ \eta_V = 2 \epsilon_V$, and one can work with ${\cal O}(1)$ value of $\alpha$.
If one tries to work with cold inflationary models with this potential, one gets power law expansion  $a \sim t^{q}$ with $q = 2/\alpha^2$. 

In warm inflation, $\eta_w = 2 \epsilon_w$. Note that while $\epsilon_V$ and $\eta_V$ are fixed quantities, since $Q$ is in general temperature and field dependent, $\epsilon_w$ and $\eta_w$ will change as inflation proceeds. 

We will find that in this model inflation automatically ends, because \eqref{endinf} is guaranteed to happen. We will use this to show that for reasonable values of $N_{\rm cmb}$, the spectrum has too much red tilt. 

\subsubsection{End of inflation}

Note that as a model of cold inflation, the potential given by Eq ($\ref{eq:oldpot}$) is incomplete in the sense that it does not have a natural end point to its evolution. But for cold inflation, even if one assumes that such a mechanism exists and does not disrupt the predictions for cosmological perturbations, this model leads to $r = 8(1-n_s)$ which, for the measured value of $n_s$ gives a value of $r$ which is ruled out.

Thus, it is worth finding out whether there is graceful exit for warm inflation in this model. For the potential of interest, we see from Eq~(\ref{Q-evo}) that 
\begin{eqnarray}
\frac1Q\frac{dQ}{dN}=-\frac{2\epsilon_V}{1+7Q},
\label{Q-evo-2}
\end{eqnarray}
which shows  $Q$ decreases with $N$. As $\epsilon_V$ is a constant in this case, $Q$ will eventually drop down to meet the condition $\epsilon_V=1+Q$ ending inflation eventually. 
This is an approximate analytical argument for the end of inflation, we have also numerically solved 
Eq (\ref{eq:phi}), Eq (\ref{eq:Hubb}) and Eq (\ref{eq:rad}) simultaneously and verified that one does achieve end of inflation in this model.

\subsubsection{Constraint on $n_s$}

Now that we know that warm inflation does end in this scenario, let us find predictions for CMB observables.
From Eq~(\ref{eq:ns}) and discussion in the beginning of \textsection \ref{sec:oldpot},  
$n_s$ (determined at $\phi_*$) is found to be:
\begin{eqnarray}
n_s=1-\frac{33}{14}\frac{\alpha^2}{1+Q_*}.
\end{eqnarray}
Let us determine the factor $\alpha^2/(1+Q_*)$. From Eq (\ref{eq:Q}), we see that 
\begin{eqnarray}
1+Q_*\sim Q_*={\tilde C}^{1/7}\frac{\alpha^{6/7}}{M_{pl}^{6/7}}V_0^{1/7}e^{-\alpha\phi_*/7M_{pl}}.
\end{eqnarray}
We can calculate $N_{\rm cmb}$ for this potential as 
\begin{eqnarray}
\frac{\alpha^2}{7}N_{\rm cmb}= {\tilde C}^{1/7}\frac{\alpha^{6/7}}{M_{pl}^{6/7}}V_0^{1/7}\left(e^{-\alpha\phi_*/7M_{pl}}-e^{-\alpha\phi_{\rm end}/7M_{pl}}\right).\nonumber\\
\end{eqnarray}
Note that because of the very fast evolution of $Q$, we cannot approximate the integral for $N_{\rm cmb}$ as the integrand times the field range. We will have more to say about it the next sub-subsection. 

From the above two equations we see that 
\begin{eqnarray}
1+Q_*=\frac{\alpha^2}{7}N_{\rm cmb}+{\tilde C}^{1/7}\frac{\alpha^{6/7}}{M_{pl}^{6/7}}V_0^{1/7}e^{-\alpha\phi_{\rm end}/7M_{pl}}.
\end{eqnarray}
One can calculated $\phi_{\rm end}$ for this potential from Eq~(\ref{phi-end}) as 
\begin{eqnarray}
{\tilde C}^{1/7}\frac{\alpha^{6/7}}{M_{pl}^{6/7}}V_0^{1/7}e^{-\alpha\phi_{\rm end}/7M_{pl}}=\frac{\alpha^2}{2}.
\end{eqnarray}
Thus we get
\begin{eqnarray}
\frac{\alpha^2}{1+Q_*}=\frac{2}{1+\frac27N_{\rm cmb}},
\end{eqnarray}
yielding 
\begin{eqnarray}
n_s=1-\,\frac{33}{7+2N_{\rm cmb}}.
\end{eqnarray}
This gives $n_s=0.74$ (0.69) for $N_{cmb}=60$ (50) respectively. 

In the above argument, we assumed that end of inflation is realized via \eqref{endinf}. This leaves open the sliver of a possibility that if we could end warm inflation using some other mechanism at some other point in field space, perhaps we could bypass the problem of an overly red-tilted spectrum.  While this is perhaps difficult to disprove, note that in this kind of potential $\phi$ rolls from smaller values to larger values. Thus,  $\phi_*<\phi_{\rm end}$. In fact it turns out in many cases (see eg., next sub-subsection)  $e^{-\alpha\phi_*}\gg e^{-\alpha\phi_{\rm end}}$, and therefore we see that the dependence on $\phi_{\rm end}$ is weak: 
\begin{eqnarray}
\frac{\alpha^2}{7}N_{\rm cmb}\sim {\tilde C}^{1/7}\frac{\alpha^{6/7}}{M_{pl}^{6/7}}V_0^{1/7}e^{-\alpha\phi_*/7M_{pl}}=1+Q_*,
\end{eqnarray}
yielding 
\begin{eqnarray}
n_s\sim1-\frac{33}{2N_{\rm cmb}}.
\end{eqnarray}
This gives $n_s=0.725$ (0.67) for $N_{cmb}=60$ (50) respectively. This again shows that the scenario yields way too much red tilted spectrum than is observationally allowed, irrespective of the details of how and where inflation ends.

Let us try to understand this from a different perspective. In cold inflation $n_s=1-6\epsilon_V+2\eta_V$. For this potential $\eta_V=2\epsilon_V$. This gives $n_s=1-2\epsilon_V$, and thus $n_s<1$ (red tilted). In warm inflation, if $\Upsilon$ is proportional to some power of $T$ then it makes the spectrum even more red tilted than in the case of cold inflation for such potentials. That is why, the potential chosen in ref \cite{Berghaus:2019whh}, which gives blue tilted spectrum ($n_s>1$) in cold inflation, leads to the observed spectral index for this scenario in warm inflation. 

Thus we require a form of potential which, in general, yields a blue tilted spectrum in cold inflationary scenario, so that both the effects can nullify each other providing a scenario compatible with observations. We will consider such a scenario in the next subsection.

\subsubsection{Field excursion}

Before closing our discussion about the potential given by Eq (\ref{eq:oldpot}), let us also note that, for ${\cal O}(1)$ values of $\epsilon_V$ (as required by refined dS Swampland conjecture), the field excursion required for achieving 40-60 e-foldings of inflation is super-Planckian, which violates Swampland distance conjecture. 
This can be seen as follows: the field excursion of the inflaton during inflation is given by
 \begin{equation}
 \frac{\Delta \phi}{M_{pl}} = \int_{\phi_*}^{\phi_{\rm end}} dN \left( \frac{-V'}{V} \right) \frac{M_{pl}}{1+Q} = \int \frac{\alpha ~dN}{1+Q} \; ,
 \end{equation}
 now, we can use Eq (\ref{Q-evo-2}) and the fact that $2 \epsilon_V = \alpha^2$ to write 
 \begin{equation}
 dN = \frac{-1}{\alpha^2} \left( \frac{1+7Q}{Q} \right) dQ \; ,
 \end{equation}
this implies that
 \begin{equation}
 \frac{\Delta \phi}{M_{pl}} = \frac{-1}{\alpha} \left[ \ln \left( \frac{Q_{\rm end}}{Q_*} \right) + 6 \ln \left( \frac{1+Q_{\rm end}}{1+Q_*} \right) \right] \; .
 \end{equation}
Now, one can use the fact that $1+ Q_{\rm end} = \alpha^2/2$ and $1+ Q_{*} = (\alpha^2/7) N_{cmb} + \alpha^2/2$, 
to find that 
 \begin{eqnarray} \label{Eq:field-excursion}
 \frac{\Delta \phi}{M_{pl}} &=& \frac{-1}{\alpha} \Bigl[ \ln \left( \frac{\alpha^2/2 - 1}{(\alpha^2/7) N_{cmb} + \alpha^2/2 - 1} \right) \nonumber \\
 &+& 6 \ln \left( \frac{1}{1+\frac{2N_{cmb}}{7}} \right) \Bigr] \; .
 \end{eqnarray}
For $\alpha < \sqrt{2}$, the argument of the first log becomes negative, so we consider values of $\alpha$ bigger than $\sqrt{2}$, the results are shown in Fig (\ref{Deltaphi}). It is seen that the field excursion is super-Planckian for $\alpha \sim {\cal O}(1)$, thus, this potential will not be consistent with the Swampland distance conjecture.
\begin{figure}
  \includegraphics[width = 0.48\textwidth]{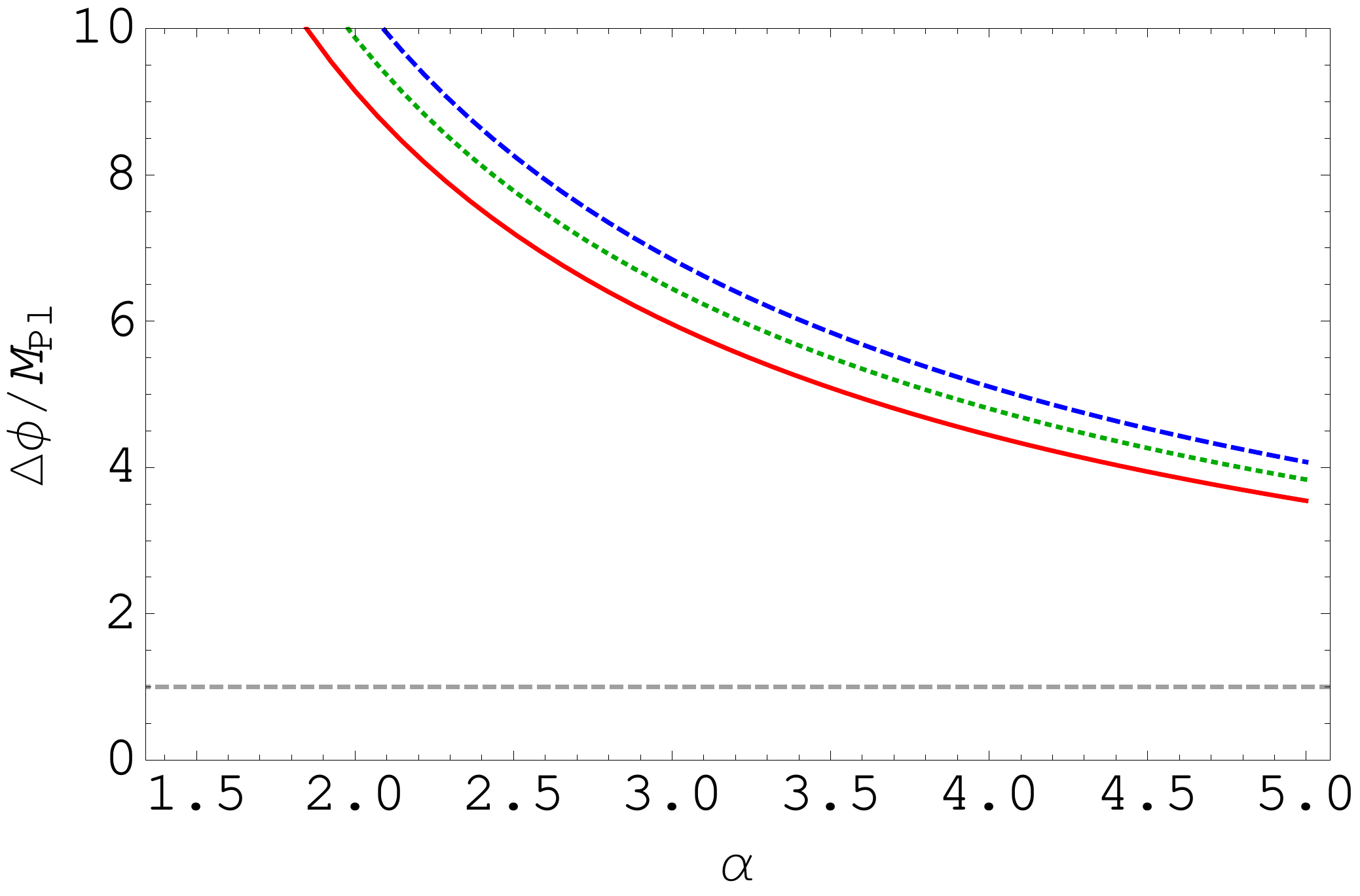}
  \caption
 {This plot shows how the field excursion found from Eq (\ref{Eq:field-excursion}) changes with $\alpha$ for $N_{cmb} = 40$ (solid, red curve), $N_{cmb} = 50$ (dotted, green curve) and $N_{cmb} = 60$ (dashed, blue curve). The dashed grey horizontal line is field excursion of $1~ M_{pl}$. 
 }
  \label{Deltaphi}
\end{figure}

We have thus found that, while the potential given by Eq (\ref{eq:oldpot}) leads to warm inflation with graceful exit, the corresponding value of scalar spectral index $n_s$ can not match with observations and the inflaton field excursion is trans-Planckian.

The fact that $Q$ varies a lot during the evolution plays a key role in the above discussion. In particular, approximating the integral by the integrand times the integration range, leads to incorrect results. Note that in this aspect, this simplest swampland runaway potential is different both from the examples in \cite{Berghaus:2019whh} as well as from our own example in the next subsection.

\subsection{Potential of the form $V = V_0  \left[1 + e^{-\alpha \frac{\phi}{M_{pl}}} \right]$}

In this section, we work with a potential similar to the one given by Eq (\ref{eq:oldpot}), but with a constant added to it: thus, consider a potential of the form
\begin{equation} \label{eqpotnew}
V = V_0  \left[1 + e^{-\alpha \frac{\phi}{M_{pl}}} \right] \; .
\end{equation}
Before we analyse the conditions under which this potential is consistent with swampland conjecture and with observations, let us see what happens if we were considering a more general potential:
\begin{equation}
V = V_0 \left[ \gamma + \beta e^{-\alpha \frac{\phi}{M_{pl}}} \right] \; ,
\end{equation}
then, we could have written it as 
\begin{equation}
V = V_0 \gamma \left[ 1 + \frac{\beta}{\gamma} e^{-\alpha \frac{\phi}{M_{pl}}} \right]  = V_0 \gamma \left[ 1 + e^{\ln \frac{\beta} {\gamma}} e^{-\alpha \frac{\phi}{M_{pl}}} \right]  \; ,
\end{equation}
which is the same as
\begin{equation}
V = {\tilde V}_0 \left[ 1 +  e^{-\alpha \frac{\tilde \phi}{M_{pl}}} \right] \; ,
\end{equation}
for redefined field and parameter values: thus, Eq (\ref{eqpotnew}) captures a large class of possibilities.

\subsubsection {The slow roll parameters and swampland conditions}

We now wish to find the conditions under which the potential given by Eq (\ref{eqpotnew}) is consistent with swampland constraint. For this potential we have 
\begin{eqnarray} \label{eq:srnew}
\epsilon_V&=&\frac{\alpha^2}{2}\left(\frac{e^{-\alpha\phi/M_{Pl}}}{1+e^{-\alpha\phi/M_{Pl}}}\right)^2 \; , \\ 
\eta_V&=&\alpha^2 \left( \frac{e^{-\alpha\phi/M_{Pl}}}{1+e^{-\alpha\phi/M_{Pl}}}\right) \; ,
\end{eqnarray}
which implies that, for this potential, the following relation must always hold good
\begin{equation}
\epsilon_V = \frac{\eta_V^2}{2 \alpha^2} \; .
\end{equation}
Furthermore, when $\phi \rightarrow -\infty$ (i.e. when $\phi$ takes large negative values), we shall have 
\begin{eqnarray} 
\epsilon_V&=&\frac{\alpha^2}{2} \; ,\\
\eta_V&=&\alpha^2 \; .
\end{eqnarray}
On the other hand, when $\phi \rightarrow + \infty$, $e^{-\alpha\phi/M_{Pl}} \ll 1$ and we will be in the regime in which $\epsilon_V \ll 1$ and $\eta_V \ll 1$ while $\eta_V \gg \epsilon_V$. Notice that for all real values of $\phi$, $\eta_V > 0$, so even if we find a range of field values such that $\epsilon_V \ll 1$ and $\eta_V \sim {\cal O}(1)$ we will not be in agreement with the refined dS Swampland bound, Eq (\ref{eq:ds-swampland-condition}). This means that the potential give by Eq (\ref{eqpotnew}) will satisfy swampland conditions only for some region in field space. For any chosen value of $\eta_V$, one has
\begin{equation} \label{eq:phi-eta}
\frac{\phi}{M_{pl}} = \frac1\alpha \ln \left( \frac{\alpha^2}{\eta_V} - 1\right) \; ,
\end{equation}
equivalently, for any chosen value of $\epsilon_V$, we have 
\begin{equation}
\frac{\phi}{M_{pl}} = \frac1\alpha \ln \left( \frac{\alpha}{\sqrt{2 \epsilon_V}} - 1\right) \; .
\end{equation}
Thus, if we require $\epsilon_V$ to be greater than some ${\cal O}(1)$ number, the above equation will provide a corresponding maximum possible field value, say $\phi_{\rm max}$ (which depends on $\alpha$).
For any choice of $\alpha$, for the entire range of values of $\phi$ such that $ - \infty < \phi < \phi_{\rm max}$, refined dS swampland conditions are satisfied.
 
The important question is, for this potential, in cold inflation can one get a blue tilted spectrum such that it is good potential to try in warm inflation model with $\Upsilon\propto T^3$? If this is true, then it may yield the correct red tilted spectrum, this requires us to choose the parameters.  

\subsubsection{Limits on parameters}

When dealing with potential of this form, the free parameters are $V_0$, $\alpha$ as well as ${\tilde g}_*$, $N_{\rm cmb}$ and ${\tilde c}$. 

Let us recall that the inflaton in this model has an axionic coupling to a non-Abelian gauge theory and the sphaleron transitions between gauge vacua, existing at sufficiently high temperatures, provide the friction necessary for warm inflation. If the corresponding non-Abelian gauge theory has gauge group $SU(3)$, there will be 8 gauge bosons, each of which will contribute two relativistic degrees of freedom, and so, including the inflaton itself, there will be 17 relativistic degrees of freedom.
In most of the rest of this section, we shall present the results for the case for which $g_* = 17$ and find what happens when we change the values of the other variables.

Recall the definition of $\tilde c$, Eq ($\ref{def:ctilde}$), which is
\begin{equation}
 M_{pl}^2 ~{\tilde c} = \frac{\kappa \alpha_g^5}{(f/M_{pl})^2} \; ,
\end{equation}
 with $\alpha_g = g^2/4\pi$. We shall take $\kappa \sim 10^2$ (see Eq (\ref{eq:upsi})), the gauge coupling $g \sim 10^{-1}$ (this implies $\alpha_g \sim 10^{-3}$). Thus,
\begin{equation}
 M_{pl}^2 ~{\tilde c} = \frac{10^{-13}}{(f/M_{pl})^2} \; .
\end{equation}
In addition, we'd like to ensure that $f < V_*^{1/4}$. 
To understand this, recall that the potential that is responsible for inflation is a potential generated by UV effects, which softly breaks the shift symmetry without causing too much backreaction \cite{Berghaus:2019whh}. The scale $f$ determines the discrete shift symmetry of the unbroken IR potential. We thus expect the scale associated with IR potential to be below the scale associated with UV potential. 
In order to be consistent with TCC, we shall be interested in the case in which $V_*^{1/4} \lesssim 10^{-10} M_{pl}$. Since $f < V_*^{1/4}$, this will require $f/M_{pl} \lesssim 10^{-11}$ and hence, 
\begin{equation}
M_{pl}^2 ~ {\tilde c} \gtrsim 10^{9} \; .
\end{equation}
Assuming that $V_0$ is of the same order of magnitude as $V_*$, we now choose,
 \begin{equation}
 V_0 = 10^{-41} M_{pl}^4 \; .
 \end{equation}
 Similarly, one could let $N_{\rm cmb} \in \{ 40, 50, 60 \}$, we shall mostly show results for the case $N_{cmb} = 60$.

\subsubsection{End of inflation}

The condition for warm inflation is that the slow-roll parameters appropriate for warm inflation, given by Eq (\ref{eq:epsw}) and Eq (\ref{eq:etaw}), should be small as compared to 1. This means that $Q \gg \epsilon_V$ and $Q \gg \eta_V$, while at the end of inflation, $Q$ must fall below $\epsilon_V$ and $\eta_V$.
We will show here that for this potential both $Q$ and $\eta_V$ decrease with $N$. If inflation has to end in this scenario $Q$ has to decrease with a faster rate than $\eta_V$ to meet the condition $\eta_V=1+Q$ to end inflation. 

From Eq~(\ref{Q-evo}) we get the rate at which $Q$ evolves:
\begin{eqnarray}
\frac{d\ln Q}{dN} = \frac{10\epsilon_V-6\eta_V}{1+7Q} =  \frac{5 \left( \frac{\eta_V}{\alpha} \right)^2 - 6\eta_V}{1+7Q}.
\end{eqnarray}
Thus, $Q$ decreases when $\eta_V<1.2\alpha^2$, which is always the case, as $\eta_V$ saturates to $\alpha^2$ at large negative field values.
Similarly, from the expression of $\eta_V$ we get,
\begin{eqnarray}
\frac{d\eta_V}{dN}&=&-\frac{M_{pl}^4}{1+Q}\left[\frac{V'''}{V}\frac{V'}{V}-\frac{V''}{V}\frac{V'^2}{V^2}\right] \; , \nonumber\\
&=&-\frac{M_{pl}^4}{1+Q}\left[\frac{\alpha^4}{M_{pl}^4}\frac{e^{-2\alpha\phi/M_{pl}}}{(1+e^{-\alpha\phi/M_{pl}})^2}-\frac{V''}{V}\frac{V'^2}{V^2}\right] \; , \nonumber \\
&=&-\frac{1}{1+Q}\left(\eta_V^2-2\eta_V\epsilon_V\right) \; , \nonumber \\
&=&  -\frac{\eta_V}{1+Q} \left( \eta_V -  \left( \frac{\eta_V}{\alpha} \right)^2 \right) \; ,
\end{eqnarray}
and $\eta_V$ decreases when $\eta_V<\alpha^2$ (which, again, always holds true) with a rate:
\begin{eqnarray}
\frac{d\ln \eta_V}{dN} = -\frac{1}{1+Q} \left( \eta_V -  \left( \frac{\eta_V}{\alpha} \right)^2 \right) \; .
\end{eqnarray}
Thus, in the regime $Q\gg 1$, the ratio of rates at which $Q$ and $\eta_V$ decrease in the strongly dissipative regime is then given by 
\begin{eqnarray}
\sigma \equiv \frac{d\ln Q}{dN}/\frac{d\ln \eta_V}{dN} = \frac{ 6 - 5 \left( \frac{\eta_V}{\alpha^2} \right) }{7 \left( 1 -  \left( \frac{\eta_V}{\alpha^2} \right) \right)} \; ,
\end{eqnarray}
Fig (\ref{sigma}) shows how the quantity $\sigma$ depends on the ratio $\frac{\eta_V}{\alpha^2}$. From the above expression
we learn that if $\eta_V < \alpha^2/2$, then $\sigma < 1$ while if $ \alpha^2 > \eta_V > \alpha^2/2$, then $\sigma > 1$.
\begin{figure}
  \includegraphics[width = 0.48\textwidth]{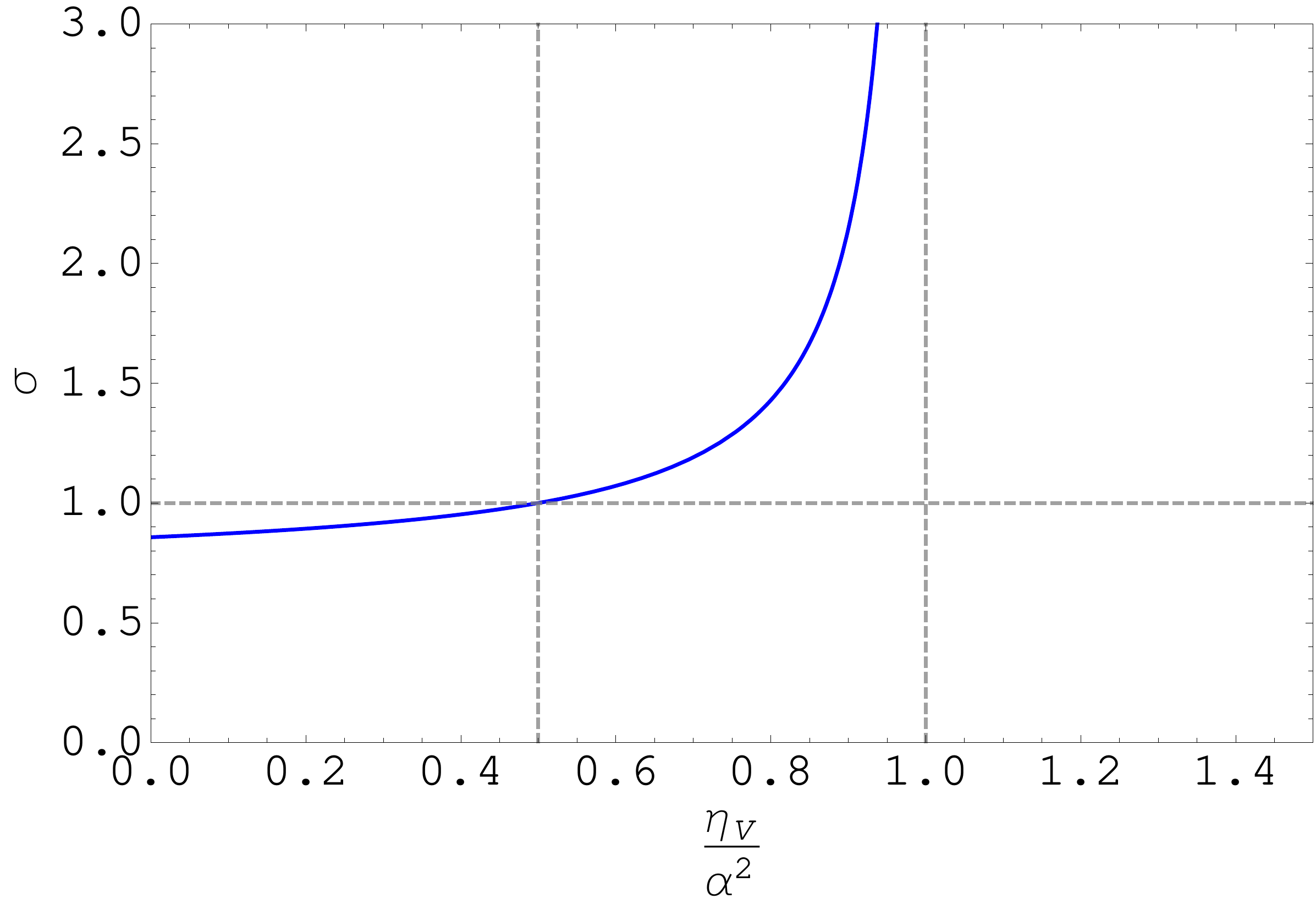}
  \caption
 {This plot shows how the ratio $\sigma$ depends on the ratio $\frac{\eta_V}{\alpha^2}$. Reality of $\phi$ in Eq (\ref{eq:phi-eta}) tells us that this ratio is required to be less than 1. For inflation to end, it is necessary that this ratio be greater than 0.5 for most of the field range, but this condition is not sufficient. 
 }
  \label{sigma}
\end{figure}

For inflation to end, it is necessary that, as we increase $N$, the quantity $Q$, which starts off being large (for warm inflation in strongly dissipative regime), must fall off faster than $\eta$ falls with increasing $N$. Thus, for inflation to end, it is necessary that $\sigma$ stays greater than 1 for a large range of $N$, but this may not be sufficient. 
This has been illustrated in Fig (\ref{rate}) for a specific example about which we will have more to say in the next subsection. One can see that, in the beginning, $1+Q$ (the red curve) falls faster than $\eta$ (the blue curve), but eventually, the curves fall equally fast and $1+Q$ doesn't fall below $\eta$.
Similarly, when $\eta_V < \alpha^2/2$ for the entire range of $\phi$ values, then $\sigma < 1$ and $Q$ decreases with a slightly slower rate than $\eta_V$, and so, in this case, it is guaranteed that there will be no end of inflation
because with a slower rate of decrease, $Q$ will never catch up with $\eta_V$ to end inflation.

\begin{figure}
  \includegraphics[width = 0.48\textwidth]{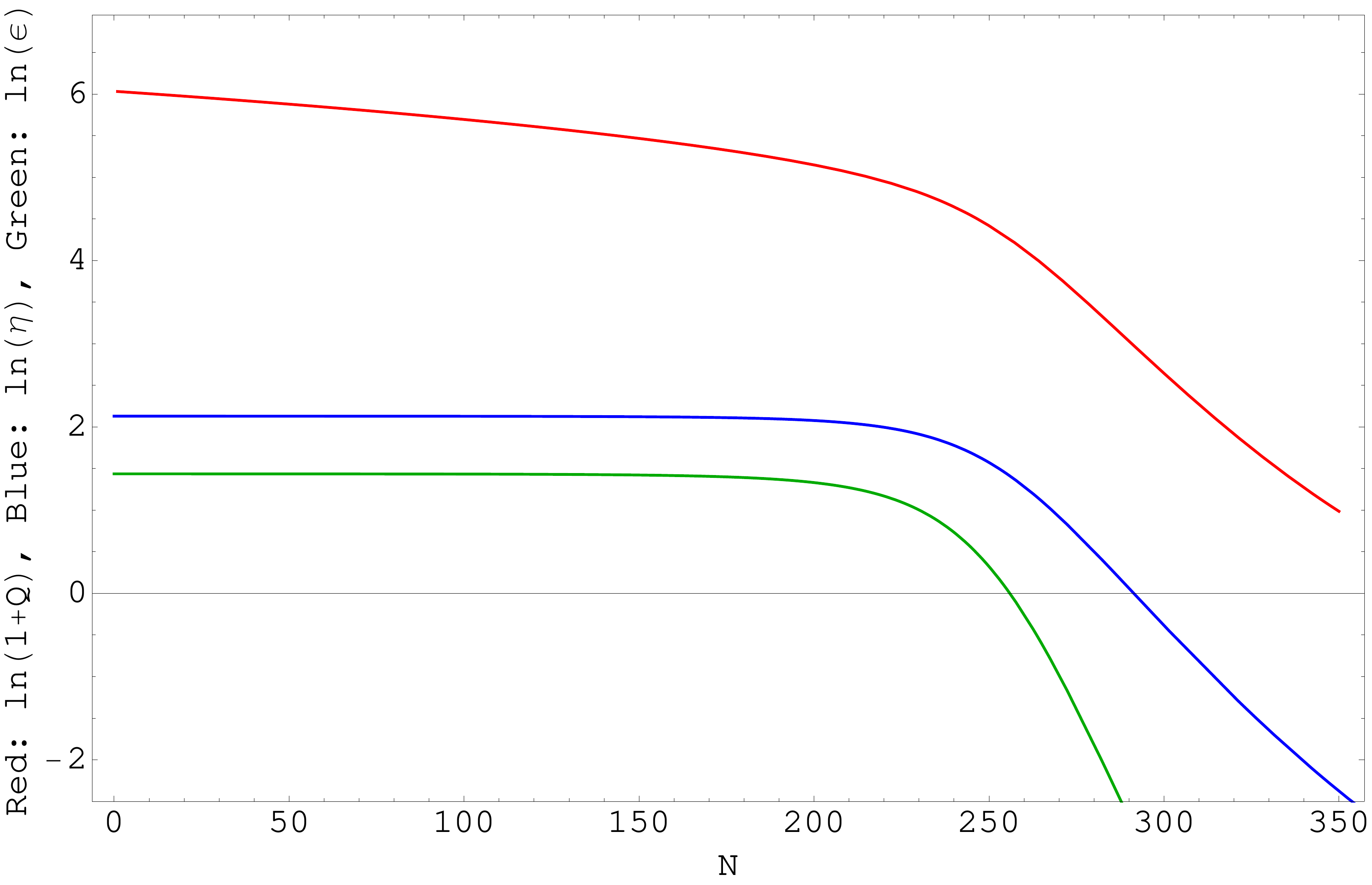}
  \caption
 {This plot illustrates the fact that, as inflation proceeds, briefly $1+Q$ can fall faster w.r.t. $N$ (the number of e-foldings of inflation), than $\eta_V$ does, but if this condition is not satisfied for sufficiently long duration, inflation won't end.
 }
  \label{rate}
\end{figure}

Thus, in this class of models, one might need an additional mechanism to end inflation for such a scenario. In the rest of this paper, we shall assume that such a mechanism can be found and restrict our attention to the swampland consistent predictions for inflation.

\subsubsection{The parameters}

The fact that inflation doesn't end in this scenario has further implications, such as the background dynamics does not set $\phi_{\rm end}$ and subsequently the value of $\phi_*$. In other words, we can have the liberty to choose $\phi_*$ which can yield the observable parameters related to the inflationary perturbations, as well as help us meet the Swampland conjectures.

Thus, we will first choose $\phi_*$ accordingly such that one can have $\epsilon_V$ greater than unity at the beginning of inflation and doesn't fall much below 0.5 after 60 $e-$folds of inflation, in order to meet the dS swampland conjecture.
We'd like ${\phi_*}$ to be less than ${M_{pl}}$ (and also less than $\phi_{\rm max}$), for a generic value of $\alpha$, we can find $\tilde c$ such that (a) these requirements get satisfied, (b) $A_s$ takes up the correct observable value. We quote the results in Table~\ref{table:cases1}

For the chosen value of $g_*$ and $V_0$ and $\alpha$ (the entry in the first row), we first set $\phi_*$ (entries in row 4) close to $\phi_{\rm max}$ such that $\epsilon_{V_*}\sim1$ is ensured. We then choose $\tilde c$ (entries in row 2) accordingly which will allow $A_s$ (entries in row 11) to be in the observed range. We employ Eq (\ref{eq:srnew}) to find the entries in row 5 and row 6, Eqs (\ref{eq:QandT}) to find entries in row 7 and row 9. We then use Eq (\ref{eq:ns}) and Eq (\ref{eq:r}) to find the entries in row 12 and 13 respectively.

Here, we note that (a) the values of $H_*$ in the table are well above the value of $H$ during big bang nucleosynthesis as we want, (b) $T_*$ is greater than $H_*$ as required by warm inflation, and (c) the tensor to scalar ratio, $r$ is unobservably small for the entire range of values. Most importantly we can observe that the $n_s$ matches best with the observation around $\alpha=3$. 
To explain this feature, we note that, for this potential:
\begin{eqnarray}
n_s=1+\frac{3}{7(1+Q_*)}(27\epsilon_{V_*}-19\sqrt{2\epsilon_{V_*}}\alpha).
\end{eqnarray}
As  $\epsilon_{V_*}\sim1$ for all the cases and we see from the table that $Q_*\sim 600$ for almost all the cases, putting these values in the above equation yields  $n_s\sim0.964$ when $\alpha\sim2.88$. However, to answer why $Q$ turns out to be of the order 600 requires a closure look at how we have arrived at the parameters furnished in the table. 
As we mentioned above, we are varying $\tilde c$  in order to get correct observed values of $A_s$. Inserting equations (\ref{eq:QandT}) into the equation for $A_s$ given in Eq (\ref{eq:As}), one can solve for $\tilde c$ for the observed value of $A_s$, and will get only one positive real root:
\begin{eqnarray}
M_{pl}^2\tilde c \sim 6 \tilde g_*^{71/90}M_{pl}^{1/5} \frac{V^{9/10}}{V'^{19/15}}.
\end{eqnarray}
Putting this value back into the expression of $Q$ given in Eqs (\ref{eq:QandT}), one gets
\begin{eqnarray}
Q_*&=&\left(\frac32\right)^{2/7}\frac{\tilde g_*^{1/45}}{V_*^{1/15}}\left(2\epsilon_{V_*}M_{pl}^4\right)^{1/15}\nonumber\\
&\sim& \left(\frac32\right)^{2/7}\frac{\tilde g_*^{1/45}}{V_0^{1/15}}\left(2\epsilon_{V_*}M_{pl}^4\right)^{1/15}.
\end{eqnarray}
Hence, for $\epsilon_{V_*}\sim 1$ one obtains $Q_*\sim 661$.

\begin{table}[h] 
\begin{tabular}{l l c c c }
&&&&\\
\hline
\hline
&&&&\\
Sr. & Qty & Case 1 &  Case 2 &  Case 3 \\
No. &&&&\\
\hline
\hline
&&&&\\
1 & $\alpha$ & $2$ & $3$ & $5$   \\
&&&&\\
2 & $M_{pl}^2~{\tilde c}$ & $9.70 \times 10^{15}$ & $1.27\times10^{16}$ & $1.41 \times 10^{16}$  \\
& &  & &  \\
3 & $\frac{\phi_{\rm max}}{M_{pl}}$ & $-0.44$ & $0.038$ & $0.186$   \\
&&&&\\
4 & $\frac{\phi_*}{M_{pl}}$ & $-0.5$ & $0.03$ & $0.18$   \\
&&&&\\
5 & $\eta_{V_*}$ & $2.92$ & $4.29$ & $7.23$   \\
&&&&\\
6 & $\epsilon_{V_*}$ & $1.07$ & $1.03$ & $1.04$   \\
&&&&\\
7 & $Q_*$ & $611.5$ & $637.4$ & $651.4$   \\
&&&& \\
8 & $\frac{V^{1/4}_*}{M_{pl}}$ & $7.8 \times 10^{-11}$ & $6.6\times10^{-11}$ & $6.1 \times 10^{-11}$   \\
&&&&\\
9 & $\frac{T_*}{M_{pl}}$ & $8.73 \times 10^{-12}$ & $7.24\times10^{-12}$ & $6.7 \times 10^{-12}$  \\
&&&&\\
10 & $\frac{H_*}{M_{pl}}$ & $3.52 \times 10^{-21}$ & $2.52\times10^{-21}$ & $ 2.16 \times 10^{-21}$ \\
&&&&\\
11 & $A_{s}$ & $2.09\times10^{-9}$ & $2.09\times10^{-9}$ & $2.08\times10^{-9}$  \\
&&&&\\
12 & $n_{s}$ & $0.981$ & $0.964$ & $0.928$   \\
&&&&\\
13 & $r$ & $7.3 \times 10^{-31}$ &$3.93\times10^{-31}$ & $2.9 \times 10^{-31}$  \\
&&&&\\
14 & $\frac{\phi_{f}}{M_{pl}}$ & $-0.35$ & $0.16$ & $0.31$   \\
&&&&\\
15 & $\frac{\Delta \phi}{M_{pl}}$ & $0.15$ &$0.13$ & $0.13$   \\
&&&& \\
16 & $\eta_{V_{f}}$ & $2.68$ &$3.41$ & $4.38$   \\
&&&&\\
17 & $\epsilon_{V_{f}}$ & $0.90$ & $0.65$& $0.38$   \\
&&&&\\
18 & $Q_f$ & $551.57$ &$510.29$ & $414.84$   \\
&&&&\\
19 & $\left.\frac{\rho_R}{V}\right|_f$ & $7\times10^{-4}$ &$6\times10^{-4}$ & $4\times10^{-4}$   \\
&&&&\\
 \hline 
\end{tabular}
\caption{The first two and the fourth rows have freely chosen values of the parameters ($\alpha$, $M_{pl}^2~{\tilde c}$ and $\phi_*$) 
(with $g_*$ chosen to be $17$, $V_0$ chosen to be $10^{-41} ~M_{pl}^4$ and $N_{\rm cmb}$ chosen to be $60$). The rest of the rows contain values of the parameters derived from the chosen values of $\alpha$, $V_0$, $M_{pl}^2~{\tilde c}$, $g_*$  and $N_{\rm cmb}$.
}
\label{table:cases1}
\end{table}

{\it Calculating $\phi_f$}: It is crucial to figure out at what $\phi$ value we need to end inflation in order to get 50-60 $e$-foldings of infation, we call this value of field $\phi_f$. 
It seems difficult to analytically figure out what $\phi_f$ would be from the expression of $e$-foldings with this form of the potential. In that case, one needs to numerically solve for it.  We see that 
\begin{eqnarray}
\frac{\dot\phi}{H}=\frac{d\phi}{dN}&=&-\frac{M_{Pl}^2}{1+Q}\frac{V'}{V},\nonumber\\
&=&\frac{M_{Pl}^2}{\tilde C^{1/7}}\left(\frac{\alpha}{M_{pl}V_0}\right)^{1/7}\frac{e^{-\alpha\phi/7M_{Pl}}}{(1+e^{-\alpha\phi/M_{Pl}})^{2/7}}.\nonumber\\
\end{eqnarray}
One then can start with $\phi=\phi_*$, and increase $\Delta N=1$ in each step from $N=0$, and go up to $N=60$ to find $\phi_f$ quoted in row 14.
This value of $\phi_f$ has then been used to find the values of the quantities in row 15, 16, 17, 18 and 19 of Table \ref{table:cases1}. Here we note that $Q$ evolves very slowly during the course of inflation in this scenario, as has been pointed out at the end of previous subsection. We also note that $\rho_R$ by the end of 60 $e$-foldings is much less than $\rho(\phi)$, which also carries a signature that Warm inflation does not end in such a scenario.

\subsubsection{Comparison with exact numerical evolution}

For numerical solution of Eqs (\ref{eq:phi},\ref{eq:Hubb},\ref{eq:rad}), one could choose the values of the free parameters: e.g. we choose $\alpha = 3$, $V_0 = 10^{-41} ~M_{pl}^4$, $M_{pl}^2~{\tilde c} = 1.2703 \times 10^{16}$, $g_* = 17$. We can then begin with some initial field value $\phi_{\rm ini} < \phi_*$. Note that when $\phi_{\rm ini}$ is chosen to be sufficiently smaller than $\phi_*$, the exact chosen initial values of $\dot \phi$ and $T$ are unimportant. Under such conditions, one could simply choose initial values of $\dot \phi$ and $T$ to be zero.

With such a choice of the initial conditions (and the free parameters), one can evolve Eqs (\ref{eq:phi},\ref{eq:Hubb},\ref{eq:rad}) numerically and obtain the values of all the quantities in table \ref{table:cases1}. The values of the quantities found from this more accurate procedure are found to be very close to the values presented in the table. 
E.g. we found that 
$\eta_* = 4.297$, $\epsilon_* = 1.026$, $Q_* = 636.643$, $V_*^{1/4} = 6.459 \times 10^{-11} ~M_{pl}$, $T_* = 7.243 \times 10^{-12}$, $H_* = 2.527 \times10^{-21}$ and 
$\eta_f = 3.413$, $\epsilon_f = 0.647$, $Q_f = 509.732$, $V_f^{1/4} = 5.877 \times 10^{-11} ~M_{pl}$, $T_f = 6.535 \times 10^{-12}$, $H_f = 2.318 \times10^{-21}$: these values should be compared with the entries in case 2 of table \ref{table:cases1}.

This verifies the results of the previous section. Furthermore, one finds that, even if one starts with zero temperature of the thermal bath, very soon, conditions for warm inflation get established (as was argued in \cite{Berghaus:2019whh}).

\section{Discussion}

In this work we studied the warm inflationary predictions of the simple scalar potential given by Eq (\ref{eqpotnew}). From the second column of table (\ref{table:cases1}), it is clear that one can choose the parameters $\alpha$ and $V_0$ in the potential, as well as $\tilde c$, such that 
\begin{itemize}
\item CMB observables ($n_s$ and $A_s$) take up their observed values, 
\item $Q_* \gg 1$ so that warm inflation in the strongly dissipative regime gets realised, 
\item the corresponding value of $V_*$ is consistent with the trans-Planckian censorship conjecture,
\item the field excursion is sub-Planckian as required by swampland distance conjecture, and, 
\item for the relevant range of fields, the potential is consistent with the dS swampland conjectures. 
\end{itemize}
This demonstrates that Minimal Warm Inflation \cite{Berghaus:2019whh} provides a viable realization of inflation that can simultaneously meet observational data while satisfying all of the relevant swampland constraints in an extremely simple model. 

There are a few observations here that are worth making. As we discussed in the text, the simplest runaway ptentials that satisfy the dS swampland conjectures produce too much red tilt in the spectrum and also marginally violate the swampland distance bound. We have therefore looked for and found a ``next-to-simplest" model (\ref{eqpotnew}) that is swampland-viable\footnote{Because of the swampland distance bound, it is only meanignful to discuss the validity of the swampland conditions in {\em some} ${\cal O}(1)$ range in the field space of an Effective Field Theory potential. Our potential satisfies this easily, as we discussed in section III.B.1.}. Remarkably, we find that now the observational and swampland constraints are very comfortably satisfied. We have also checked our semi-analytic estimates against explicit numerical evolution and found an excellent match.
  
While the fact that all the constraints could be satisfied in a very simple model is encouraging, to conclude we will point out various features of this approach that needs to be sorted out before it can qualify as a fully satisfactory model of inflation. 

\begin{itemize}
\item We have not discussed mechanisms for ending inflation or for ensuring reheating. These are necessary steps for transitioning to the Big-Bang nucleosynthesis phase and to have a complete cosmological model.
We have kept our discussion limited to a very basic and simple model: it seems likely that with a more specific model with more dynamical ingredients, these things can be arranged. 
\item In some of our discussions of the temperature dependence of the $\Upsilon$ parameter for various gauge groups, we have extrapolated results known only for  small gauge group ranks \cite{Moore:2010jd} to higher rank gauge groups. But this is a minor issue, which does not seriously affect our punchlines.
\item One of the features of the construction of \cite{ Berghaus:2019whh} is that their UV-potential breaks the shift symmetry of the axion completely. Since the axion arises as an angle field, even after symmetry-breaking and non-perturbative effects, we expect it to have a discrete shift symmetry of $2 \pi$ in suitable units. So the breaking here is an explicit breaking of the $2 \pi$ periodicity, but  it is claimed \cite{ Berghaus:2019whh} that it is soft. Questions regarding the breaking of what can be viewed as a discrete gauge symmetry have been discussed previously in a related set up in  \cite{Rick}, see also related very recent discussions in \cite{Reece}. It will  also be interesting to consider thermal backreaction questions from this perspective. We will  not have much to add to these discussions. A closely related question is that of realizing these in string theory, where a lot has been discussed about (the difficulty with) large axion field ranges  \cite{Palti:2019pca}.
\item A related point is that following \cite{Berghaus:2019whh}, we have demanded that the scale of the UV-potential be hierarchically above the axion decay constant. It will be good to have a better understanding of the two scales.
\end{itemize}

\begin{acknowledgements}
The authors would like to thank Rudnei Ramos and Kim Berghaus for useful discussions. The work of S.D. is supported by Department of Science and Technology, Government of India under the Grant Agreement number IFA13-PH-77 (INSPIRE Faculty Award).
\end{acknowledgements}

\end{document}